\documentclass[12pt]{article}
\usepackage{cite}
\usepackage{enumerate}
\usepackage{ifpdf}
\usepackage{ae}
\usepackage[T1]{fontenc}
\usepackage[ansinew]{inputenc}
\usepackage{amsmath}
\usepackage{relsize}
\usepackage{amssymb}
\usepackage{tikz}
\usepackage{graphicx}
\usepackage{color}
\definecolor{darkblue}{cmyk}{0.9,0.9,0,0}
\definecolor{darkgreen}{rgb}{0,0.55,0}
\usepackage[colorlinks=true,linkcolor=darkblue,citecolor=darkblue,urlcolor=darkblue]{hyperref}
\usepackage{epsfig}
\usepackage{epstopdf}
\usepackage{graphicx}

\makeatletter
\long\def\@makecaption#1#2{
  \vskip\abovecaptionskip
  \sbox\@tempboxa{{\captionfonts #1: #2}}
  \ifdim \wd\@tempboxa >\hsize
    {\captionfonts #1: #2\par}
  \else
    \hbox to\hsize{\hfil\box\@tempboxa\hfil}
  \fi
  \vskip\belowcaptionskip}
\makeatother

\newcommand{\beq}{\begin{equation}}
\newcommand{\eeq}{\end{equation}}
\newcommand{\beqy} {\begin{eqnarray}}
\newcommand{\eeqy} {\end{eqnarray}}
\newcommand{\bsmat}{\begin{smallmatrix}}
\newcommand{\esmat}{\end{smallmatrix}}
\newcommand{\bmat}{\begin{matrix}}
\newcommand{\emat}{\end{matrix}}

\def\({\left(}
\def\){\right)}
\def\[{\left[}
\def\]{\right]}

\def\<{\langle}
\def\>{\rangle}

\usepackage{varioref}
\usepackage{makeidx}
\makeindex

\usepackage[english]{babel}
\usepackage{caption}
\usepackage{subfig}

\usepackage{tabularx}
\begin{document}

\thispagestyle{empty}

\renewcommand{\thefootnote}{\fnsymbol{footnote}}
\setcounter{page}{1}
\setcounter{footnote}{0}
\setcounter{figure}{0}

\begin{titlepage}

\begin{center}

\vskip 2.3 cm 

\vskip 5mm

{\Large \bf Crossing symmetry and Higher spin towers}
\vskip 0.5cm

\vskip 15mm

\centerline{Luis F. Alday$^{\tau}$ and Agnese Bissi$^{\tau'}$  }
\bigskip
\centerline{\it $^{\tau}$ Mathematical Institute, University of Oxford,} 
\centerline{\it  Andrew Wiles Building, Radcliffe Observatory Quarter,}
\centerline{\it Woodstock Road, Oxford, OX2 6GG, UK}
\centerline{\it $^{\tau'}$ Center for the Fundamental Laws of Nature,}
\centerline{\it Harvard University, Cambridge, MA 02138 USA}

\end{center}

\vskip 2 cm

\begin{abstract}
\noindent We consider higher spin operators in weakly coupled gauge conformal field theories. Crossing symmetry of mixed scalar correlators relates different higher spin towers and we study the consequences for the spectrum and structure constants of higher spin operators of different twists. Constraints are obtained to all loops in perturbation theory. The large spin contributions to the structure constants can be resummed into a theory-dependent prefactor times a universal factor, whose structure of poles agrees with the one that would be obtained from a Witten diagram supergravity computation, although only crossing symmetry is assumed. Finally, our results provide an all loop expression for the double null limit of mixed correlators, which is in perfect agreement with the correlator/Wilson loop correspondence.
\end{abstract}

\end{titlepage}

\setcounter{page}{1}
\renewcommand{\thefootnote}{\arabic{footnote}}
\setcounter{footnote}{0}

\newpage

 \def\nref#1{{(\ref{#1})}}

\section{Introduction}

\begin{figure}[h!]
\centering
\includegraphics[width=4in]{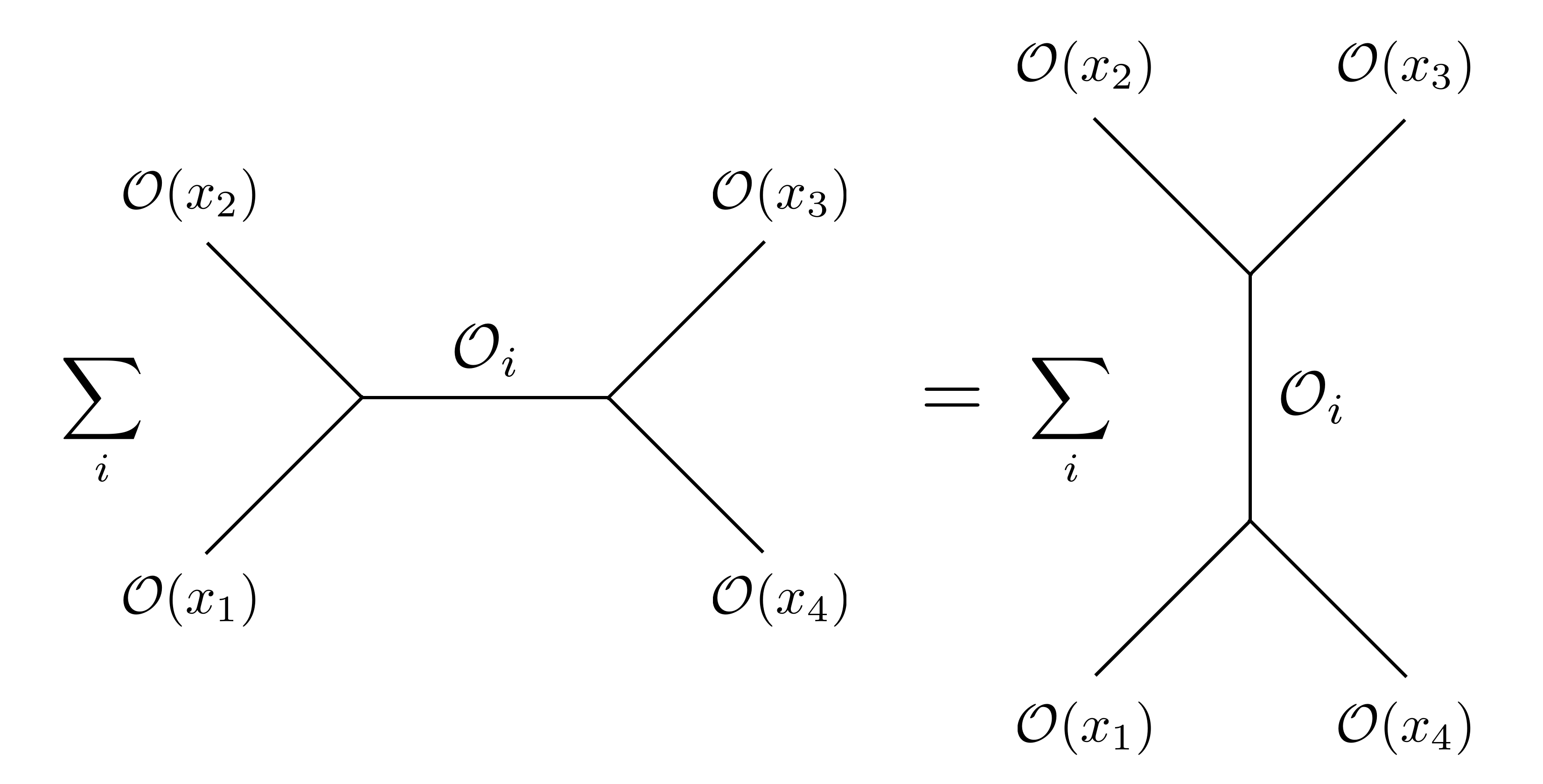}
\caption{Crossing symmetry \label{crossing}}
\end{figure}

The idea of the conformal bootstrap is that by imposing associativity of the operator product expansion (OPE) for local operators in a unitary conformal field theory (CFT) one can derive constraints for the spectrum and OPE coefficients of the theory \cite{Ferrara:1973yt,Polyakov:1974gs}. For instance, for a four-point function crossing symmetry plus the structure of the OPE  expansion schematically implies Fig.\ref{crossing}.
In a general CFT in higher dimensions, the interplay between direct and crossed channels is very complicated: a given operator on one channel will generically map to a very complicated combination of operators on  the other channel. The essence of the {\it analytic} conformal bootstrap is that for certain operators this map is much simpler. For example, each higher dimensional CFT, strongly \cite{Fitzpatrick:2012yx,Komargodski:2012ek} or weakly \cite{Alday:2015ota} coupled possesses double trace higher spin operators which map to the identity operator in the dual channel. In this paper we will concentrate on conformal weakly coupled gauge theories. These theories contain towers of higher spin operators, with small anomalous dimensions, which under crossing, and for high values of the spin, map to each other \cite{Alday:2013cwa,Alday:2015ota}:

\begin{equation}
\label{ttcrossing}
HS_\tau \leftrightarrow HS_\tau
\end{equation}
where $\tau$ denotes the twist (dimension minus the spin) of the higher spin tower. In \cite{Alday:2013cwa} we focused in a four-dimensional CFT, external operators of the form ${\cal O}^{[2]} = Tr \varphi^2$ and single-trace higher spin operators of the schematic form ${\cal O}_\ell = Tr \varphi \partial^\ell \varphi$,  with twist two and spin $\ell$. In perturbation theory these operators have a small anomalous dimension:

$$\Delta = \ell + 2 + \gamma_\ell$$
It turns out crossing symmetry is powerful enough to fix the behaviour of the anomalous dimension for large values of the spin 

$$\gamma_\ell \sim f(g) \log \ell$$
together with the OPE coefficient of ${\cal O}_\ell$  with two external operators
\begin{equation}
\label{c22}
C_{22 \ell} \sim \Gamma \left( 1- \frac{\gamma_\ell}{2} \right)
\end{equation}
Such results \cite{Alday:2013cwa,Alday:2015ota} are valid for high values of the spin but to all orders in perturbation theory! 

On \cite{Alday:2013cwa} we have focused on correlators of identical operators. In the present paper we will consider mixed correlators and show that  (\ref{ttcrossing}) is a particular example of a more general relation, in which two different towers of higher spin operators map to each other 

\begin{equation}
\label{ttpcrossing}
HS_\tau \leftrightarrow HS_{\tau'}
\end{equation}
Analysing the consequences of crossing in this case is more delicate, since higher spin operators with more than two constituent fields are highly degenerate. For $\tau >2$ we find that again crossing implies a logarithmic growing for the anomalous dimensions, but this time in the sense of a weighted average (to be defined below). Furthermore, crossing symmetry fixes the large spin behaviour of the OPE coefficient between two scalar operators and a higher spin operator. More precisely, we obtain the following universal behaviour 

\begin{equation}
\label{proposal}
C_{pq \ell} \sim \Gamma \left( \frac{\Delta_p+\Delta_q -\tau_\ell}{2} \right)
\end{equation}
for several families of operators, as will be specified below. This is a natural generalisation of (\ref{c22}). Furthermore, note that the same structure of poles would appear in a Witten diagram supergravity computation of the three-point function for operators of dimensions $\Delta_p,\Delta_q$ and $\tau_\ell$, see for instance \cite{Freedman:1998tz,Bargheer:2013faa,Minahan:2014usa}. However, our result arises from crossing symmetry alone, without assuming large $N$ or large $R-$charges, and is a all-loop result! 

Having solved for the structure constants of higher spin operators we can study the mixed correlators under consideration in the double null limit. Our results are in perfect agreement with the picture of \cite{Alday:2010zy}. 

This paper is organised as follows. In the next section we show that crossing symmetry for mixed correlators leads to relations among different higher spin towers. In section three we derive integral relations arising from crossing symmetry for mixed correlators. Although the method is very general, we apply it to a simple model for definiteness. We then show how to solve such integral relations, finding constraints, to all loops in perturbation theory, for the spectrum of higher spin operators and OPE coefficients of the theory. As an interesting application, we compute the double null limit of the correlators under consideration and compare our results to \cite{Alday:2010zy}. In section (\ref{MSYM}) we study in detail the case of weakly coupled ${\cal N}=4$ SYM. In this case the theory possesses a global $R-$symmetry and crossing symmetry acts on the representations of this symmetry as well. By considering appropriate projections we show that the equations in this case exactly reduce to the equations previously found. We end up with a discussion of our results. Several technical details, needed in the body of the paper, are deferred to the appendices. 

\section{Crossing symmetry and higher spin towers}

 \subsection{Crossing for mixed scalar operators and higher spin towers}

Let us start with a discussion of crossing relations for the most general scalar case. This was done for instance in \cite{Kos:2014bka}. For four arbitrary scalar operators we can write, see \cite{Dolan:2011dv}

\begin{equation}
\langle \phi_i(x_1)\phi_j(x_2)\phi_k(x_3)\phi_l(x_4) \rangle = \left( \frac{x_{24}^2}{x_{14}^2} \right)^{\frac{\Delta_{ij}}{2}} \left( \frac{x_{14}^2}{x_{13}^2} \right)^{\frac{\Delta_{kl}}{2}} \frac{G_{ijkl}(u,v)}{x_{12}^{\Delta_i+\Delta_j}x_{34}^{\Delta_k+\Delta_l}}
\end{equation}
where $\Delta_{ij} = \Delta_i - \Delta_j$ and we have introduce the conformal cross-ratios

\begin{equation}
u= \frac{x_{12}^2 x_{34}^2}{x_{13}^2 x_{24}^2},~~~~v= \frac{x_{14}^2 x_{23}^2}{x_{13}^2 x_{24}^2}
\end{equation}

 The full correlation function must be invariant under the exchange $(1,i) \leftrightarrow (3,k)$, which gives the crossing equation

\begin{equation}
v^{\frac{\Delta_j +\Delta_k}{2}} G_{ijkl}(u,v) = u^{\frac{\Delta_i + \Delta_j}{2}} G_{kjil}(v,u)
\end{equation}
The function $G_{ijkl}(u,v)$ can be decomposed in terms of conformal blocks as

\begin{equation}
 G_{ijkl}(u,v) = \sum_{\cal O} c_{ij {\cal O}} c_{kl {\cal O}} g_{\Delta,\ell}^{\Delta_{ij},\Delta_{kl}}(u,v)
\end{equation}
where ${\cal O}$ runs over all conformal primary operators present in the corresponding OPE and $\Delta,\ell$ denote the dimension and spin of such operators. The crossing equation implies

\begin{equation}
v^{\frac{\Delta_j +\Delta_k}{2}}  \sum_{\cal O} c_{ij {\cal O}} c_{kl {\cal O}} g_{\Delta,\ell}^{\Delta_{ij},\Delta_{kl}}(u,v) = u^{\frac{\Delta_i +\Delta_j}{2}}  \sum_{\cal O} c_{kj {\cal O}} c_{il {\cal O}} g_{\Delta,\ell}^{\Delta_{kj},\Delta_{il}}(v,u) 
\end{equation}
This is a very complicated (but powerful!) equation, as usually single operators on one channel, are mapped to infinite, complicated combinations on the other channel. The essence of the analytic bootstrap is that:

1.- Certain towers of higher spin operators map to simple operators on the other channel: either to isolated operators of low twist \cite{Komargodski:2012ek,Fitzpatrick:2012yx} or to themselves \cite{Alday:2013cwa,Alday:2015ota}.

2.- One can access this regime by considering a light-cone OPE of the four-point correlator. 

This allows to compute certain features of the spectrum and OPE coefficients of higher spin operators exactly. The simplest example arises for identical external operators with dimension $\Delta_0$. In this case the crossing relation reads

\begin{equation}
\label{crossingid}
v^{\Delta_0}  \sum_{\cal O} c_{ij {\cal O}} c_{kl {\cal O}} g_{\Delta,\ell}(u,v) = u^{\Delta_0}  \sum_{\cal O} c_{kj {\cal O}} c_{il {\cal O}} g_{\Delta,\ell}(v,u) 
\end{equation}
where we have introduced $g_{\Delta,\ell}(u,v)=g_{\Delta,\ell}^{0,0}(u,v)$. In conformal weakly coupled gauge theories there are towers of almost conserved higher spin currents of twist $\tau= \Delta - \ell= \Delta_0+ \gamma_\ell$. In such a situation the crossing relation (\ref{crossingid}) maps the large spin sector of the higher spin tower $HS_{\tau}$ to itself \footnote{Sometimes by "twist $\tau$" we will refer to the twist at zero coupling, so that the real twist of the operators is approximately $\tau$. This is commonly done when dealing with weakly coupled gauge theory. We hope this does not confuse the reader.}

\begin{equation}
HS_{\tau} \leftrightarrow HS_{\tau}
\end{equation}
This phenomenon was exploited in \cite{Alday:2013cwa,Alday:2015ota} to find information about the spectrum and OPE coefficients of higher spin operators. Considering instead mixed correlators we see that this is a special case of a more general relation

\begin{equation}
\label{network}
HS_{\tau} \leftrightarrow HS_{\tau'}
\end{equation}
Indeed, consider the contribution to $G_{ijkl}(u,v)$ from operators with twist $\tau$: $G_{ijkl}(u,v)|_\tau = u^{\tau/2} h(v)$. If $h(v)$ diverges as $v \to 0$

\begin{equation}
G_{ijkl}(u,v)|_{HS_\tau} \sim \frac{u^{\tau/2}}{v^\alpha}
\end{equation}
 then such a divergence must come from a tower of higher spin operators, of approximate twist $\tau$. Under crossing this term maps in the dual channel to a term of the form  $\frac{v^{\tau'/2}}{u^\beta}$ with
 
 \begin{equation}
v^{\frac{\Delta_j + \Delta_k}{2}} \frac{u^{\tau/2}}{v^\alpha} = u^{\frac{\Delta_i+ \Delta_j}{2}} \frac{v^{\tau'/2}}{u^\beta}
\end{equation}
Provided $\beta>0$ this must correspond to an infinite tower of higher spin operators $HS_{\tau'}$. Hence crossing leads to the relation (\ref{network}). Studying different mixed correlators will lead to different constraints involving towers of higher spin operators. Below we will study such constraints in detail, but before let us discuss the properties of different higher spin towers.

\subsection{Towers of higher spin operators}

Although the methods which we will apply in this paper are completely general, we will discuss a specific model for definiteness. Then in section \ref{MSYM} we will focus in a different model and show that the final relation has exactly the same form. Let us consider a conformal weakly coupled gauge theory in four dimensions, with a scalar field $\varphi$. The simplest gauge invariant operators are traces of such a scalar field and its derivatives. Below we will discuss the towers of higher spin operators that can arise in the light-cone OPE of scalar operators formed only by scalar fields.

\subsubsection*{Twist 2}
 These are operators of the form ${\cal O}^{(2)}_\ell = Tr \varphi \partial^\ell \varphi + \cdots$ where the derivatives are along a null direction. There is only one primary operator for each even spin and none with spin odd \footnote{In theories with other fields it is possible to form twist two operators with gauge bosons and fermions. In the presence of $R-$symmetry one can often choose a projection such that only twist-two operators made up of scalars contribute.}. We denote such a tower by $HS_2$. These operators are also called leading-twist operators and their anomalous dimension grows logarithmically with the spin:

 \begin{equation}
 \gamma_\ell^{(2)} = g \log  \ell + b(g)+ \cdots
 \end{equation} 
where we have {\it defined} the coupling constant $g$ as the coefficient in front of the logarithmic piece. All other quantities will be expressed in terms of this $g$.

\subsubsection*{Twist 3}
 These are operators of the form ${\cal O}^{(3)}_{I,\ell} =Tr \varphi^2 \partial^\ell \varphi + \cdots$. For both, spin odd and even, there is a degeneracy of primary operators, see appendix \ref{degeneracyappendix}. We denote this tower by $HS_3$. Again, their anomalous dimension grows logarithmically with the spin, but they grow along a band, as described in \cite{Belitsky:2008mg}. More precisely, for large spin
 
  \begin{equation}
g \rho_I \log \ell + \cdots \leq  ~\gamma^{(3)}_{I,\ell}~  \leq 2 g \rho_I \log \ell + \cdots 
 \end{equation} 

\subsubsection*{Twist 4}
For twist four and higher we have a new ingredient. On one hand, there are {\it single-trace} operators of the form ${\cal O}^{(4)}_{I,\ell} =Tr \varphi^3 \partial^\ell \varphi + \cdots$. There are primary operators for both, spin odd and even. Their anomalous dimension grows logarithmically with the spin, again along a band, schematically

 \begin{equation}
 \gamma^{(4)}_{I,\ell} = g \rho_I \log\ell + \cdots
 \end{equation} 
 where now $1 \leq \rho_I \leq 3$. In addition, there are double trace operators, of the form $[{\cal O}^{(2)},{\cal O}^{(2)}]_\ell ={\cal O}^{(2)}_{\ell_1} \partial^{\ell_2} {\cal O}^{(2)}_{\ell_3}$, with $\ell_1+\ell_2+\ell_3=\ell$. For large $\ell_1+\ell_3$ (a macroscopic fraction of $\ell$) their anomalous dimension again grows logarithmically with the spin. The discussion in \cite{Alday:2007mf} and a naive application of the results of \cite{Komargodski:2012ek,Fitzpatrick:2012yx} would imply an anomalous dimension of the form
 
 \begin{equation}
 \gamma^{(4)}_{DT,\ell} \sim  g \left( \log \ell_1+ \log \ell_3 \right)+  \cdots
 \end{equation} 
 While the regime considered in this paper is not in the range of validity \cite{Komargodski:2012ek,Fitzpatrick:2012yx}, the ideas of  \cite{Alday:2007mf} still apply and we expect this expansion to be true. The main difference with the single trace case is that there are operators with very large spin whose anomalous dimension does not grow logarithmically, the ones with small $\ell_1,\ell_3$. Of course, in a non-planar theory there is really no distinction between single and multi-trace operators, but the behaviour with spin will have the same features: for most operators it will  grow logarithmically but there will be some operators for which it wont.  We denote the full contribution of all operators $HS_4$ and the results of this paper will still apply. Higher twists behave in a similar way, except in general we can also have triple trace, etc. 

\section{Consequences of crossing symmetry}
Although our methods will be general, we will focus for definiteness on a specific model. We consider a conformal weakly coupled gauge theory in four dimensions and external operators of the form

\begin{equation}
{\cal A}^{L}  = Tr \varphi^L
\end{equation}
with dimension $\Delta=L$ at tree level. We will assume for simplicity that such operators are protected. Furthermore we denote by $G_{L_1L_2L_3L_4}(u,v)$ the conformal invariant part of the correlator

$$\langle {\cal A}^{L_1}(x_1){\cal A}^{L_2}(x_2){\cal A}^{L_3}(x_3){\cal A}^{L_4}(x_4) \rangle $$
We will start by considering the simplest case of four identical operators with $L=2$. This will serve to introduce some important ingredients. Then we will proceed to discuss a more general case.

\subsection{Integral relations}

\subsubsection*{Correlator $G_{2222}(u,v)$}

Let us start by the simplest case $G_{2222}(u,v)$. At tree-level we obtain
\begin{equation}
G_{2222}(u,v) = 1 + u \left(\frac{c_{11}}{v} + c_{10} \right)+ u^2 \left(\frac{c_{22}}{v^2} + \frac{c_{21}}{v}+ c_{20} \right)
\end{equation}
The constants $c_{ij}$ will in general depend on the parameters of the theory, {\it e.g.} its central charge, but their explicit form will not be relevant for us. When the coupling constant is turned on these coefficients get dressed by logarithms, and to any order in perturbation theory

\begin{equation}
c_{ij} \to c_{ij}(\log u,\log v)
\end{equation}
where the function $c_{ij}(\log u,\log v)$ is by definition the function in front of $\frac{u^i}{v^j}$ in a small $u,v$ expansion, see \cite{Alday:2015ota}. From the structure of divergences, and the powers of $u$, the contribution $c_{11}$ must come from the exchange of an infinite tower of higher spin operators with twist two, or $HS_2$. Furthermore $c_{22}$ arises solely from the tower $HS_4$,  while $c_{21}$ may receive contributions from both, twist four operators as well as descendants of twist two operators, see appendix \ref{divappendix}. What do these towers map to under crossing? Crossing symmetry implies
\begin{equation}
v^2 G_{2222}(u,v) = u^2  G_{2222}(v,u)
\end{equation}
We see $c_{22}$ gets mapped to the contribution from the identity operator. This is an example of the phenomenon studied in \cite{Komargodski:2012ek,Fitzpatrick:2012yx}. Furthermore, the contribution $c_{11}$ maps to itself, so that this corresponds to 
\begin{equation}
\label{HS22}
HS_{2} \leftrightarrow HS_{2}
\end{equation}
More precisely, crossing symmetry implies $c_{11}(\log u,\log v)=c_{11}(\log v,\log u)$. In order to study the consequences of this relation let us follow \cite{Alday:2013cwa} and consider both ways of decomposing $c_{11}$:
\begin{eqnarray}
\left. G_{2222}(u,v) \right|_{HS_2} = u \sum_{\ell}^\infty c_{22{\cal O}^{(2)}_\ell}^2u^{\gamma_\ell^{(2)}/2}  f_{coll \,(\Delta,\ell)}(v) = \frac{u}{v} c_{11}(\log u,\log v) + \cdots\\
\left. G_{2222}(v,u) \right|_{HS_2} = v \sum_{\ell}^\infty c_{22{\cal O}^{(2)}_\ell}^2 v^{\gamma_\ell^{(2)}/2} f_{coll \,(\Delta,\ell)}(u) = \frac{v}{u} c_{11}(\log v,\log u) + \cdots
\end{eqnarray}
In order to reproduce the correct divergence at tree level in either sum we require

\begin{equation}
\left( c^{(0)}_{22{\cal O}^{(2)}_\ell}\right)^2 \sim \frac{\ell^{1/2}}{4^\ell}
\end{equation}
see appendix \ref{divappendix}. Writing 

\begin{equation}
\left( c_{22{\cal O}^{(2)}_\ell}\right)^2= \left( c^{(0)}_{22{\cal O}^{(2)}_\ell}\right)^2  \hat a(\ell)
\end{equation}
the condition arising from crossing can be written as

\begin{equation}
\label{integraltwist2}
\int_0^\infty \hat a\left( \frac{x}{\sqrt{v}} \right) 2^{\gamma^{(2)}\left( \frac{x}{\sqrt{v}}\right)}u^{\gamma^{(2)}\left( \frac{x}{\sqrt{v}}\right)/2} x K_0(2x) dx = \int_0^\infty \hat a\left( \frac{x}{\sqrt{u}} \right) 2^{\gamma^{(2)}\left( \frac{x}{\sqrt{u}}\right)} v^{\gamma^{(2)}\left( \frac{x}{\sqrt{u}}\right)/2} x K_0(2x) dx 
\end{equation}
where in $\hat a(\ell)$ and $\gamma^{(2)}(\ell)$ only the contributions that do not vanish at large spin are kept. This integral relation imposes conditions on both, the spectrum and the OPE coefficients. More precisely, at large spin \cite{Alday:2013cwa}:

\begin{eqnarray}
\label{constraints2}
\gamma_\ell^{(2)} &=& g \log \ell + b(g) + \cdots\\
\hat a(\ell) &=& \kappa(g) 2^{-\gamma_\ell^{(2)}} e^{-b(g) \log \ell} \Gamma^2\left( 1- \frac{\gamma_\ell^{(2)}}{2}\right) \nonumber
\end{eqnarray}
These results are valid to all loops in perturbation theory. Namely, to all orders in $g$ in the regime $g \log \ell \ll 1$, although they resum all perturbative corrections. 

\subsubsection*{Correlator $G_{ppqq}(u,v)$}

Let us study the correlator $G_{ppqq}(u,v)$ and its permutations, the most relevant case for this paper. In this case there are two distinct crossing relations and they provide different information. Let us start by computing at zero coupling

\begin{equation}
\label{Gppqq}
G_{ppqq}(u,v) = 1 +\sum_{i=1}^{min(p,q)} \sum_{j=0}^i \frac{u^i}{v^j} c_{ij}
\end{equation}
As we turn on the coupling constant $c_{ij} \to c_{ij}(\log u,\log v)$. By crossing symmetry we then obtain \footnote{As we turn on the coupling also new higher powers of $u$ and $v$ will arise, which are not included in our formulae. Those will not be relevant for us.}

\begin{eqnarray}
\label{Gqppq}
G_{qppq}(u,v) = \frac{u^{\frac{p+q}{2}}}{v^p} G_{ppqq}(v,u) = \frac{u^{\frac{p+q}{2}}}{v^p}+ \sum_{i=1}^{min(p,q)} \sum_{j=0}^i \frac{u^{\frac{p+q}{2}-j}}{v^{p-i}}c^c_{ij}
\end{eqnarray}
where we have introduced the notation $c^c_{ij}=c_{ij}(\log v,\log u)$. Let us focus in the contributions $c_{\frac{\tau}{2} \frac{\tau}{2}}$, with $\tau=2,4,\cdots, 2\, min(p,q)$. From the point of view of the direct channel (\ref{Gppqq}) these can only arise from higher spin towers $HS_\tau$. Indeed, descendants of lower twist higher spin towers will not produce a divergent enough term,  see appendix \ref{divappendix}. From the point of view of the dual channel  (\ref{Gqppq}) these can only arise from higher spin towers $HS_{p+q-\tau}$, for the same reason. Hence, crossing relates

$$HS_\tau \leftrightarrow HS_{p+q-\tau}$$
which generalises (\ref{HS22}). Let us proceed as above and consider the OPE decomposition of $c_{\frac{\tau}{2} \frac{\tau}{2}}$ and $c^c_{\frac{\tau}{2} \frac{\tau}{2}}$:

\begin{eqnarray}
\left. G_{ppqq}(u,v) \right|_{HS_\tau} &=&  u^{\tau/2} \sum_{\ell,I} c_{pp{\cal O}^{(\tau)}_{I,\ell}} c_{qq{\cal O}^{(\tau)}_{I,\ell}}  u^{\gamma^{(\tau)}_{I,\ell}/2}f_{coll \, (\Delta_{I,\ell},\ell)}(v) \\
&=& \frac{u^{\tau/2}}{v^{\tau/2}} c_{\frac{\tau}{2} \frac{\tau}{2}}(\log u, \log v)  + \cdots \nonumber \\
\left. G_{qppq}(u,v) \right|_{HS_{p+q-\tau}} &=& u^{(p+q-\tau)/2} \sum_{\ell,I} c_{qp{\cal O}^{(p+q-\tau)}_{I,\ell}} c_{pq{\cal O}^{(p+q-\tau)}_{I,\ell}}  u^{\gamma^{(p+q-\tau)}_{I,\ell}/2}f^{(q-p,p-q)}_{coll \, (\Delta_{I,\ell},\ell)}(v) \\
&=& \frac{u^{(p+q-\tau)/2} }{v^{p-\tau/2}} c_{\frac{\tau}{2} \frac{\tau}{2}}(\log v, \log u)  + \cdots \nonumber
\end{eqnarray}
so that crossing reads

\begin{equation}
\label{crossingpqt}
\left. \sum_{\ell,I} c_{pp{\cal O}^{(\tau)}_{I,\ell}} c_{qq{\cal O}^{(\tau)}_{I,\ell}}  u^{\gamma^{(\tau)}_{I,\ell}/2}f_{coll \, (\Delta_{I,\ell},\ell)}(v) \right|_{\frac{1}{v^{\tau/2}}} = 
\left.  \sum_{\ell,I} c_{qp{\cal O}^{(p+q-\tau)}_{I,\ell}} c_{pq{\cal O}^{(p+q-\tau)}_{I,\ell}}  v^{\gamma^{(p+q-\tau)}_{I,\ell}/2}f^{(q-p,p-q)}_{coll \, (\Delta_{I,\ell},\ell)}(u) \right|_{\frac{1}{u^{p-\tau/2}}} 
\end{equation}
For a fixed $\ell$, the index $I$ labels different operators which are degenerate at tree level. Twist two is the only non-degenerate case. The next step is to convert this relation into an integral equation. This is a bit more subtle than before, as for each spin we have a degeneracy at tree-level. In order to reproduce the correct divergence at tree level we must have

\begin{eqnarray}
& &\sum_I  c^{(0)}_{pp{\cal O}^{(\tau)}_{I,\ell}} c^{(0)}_{qq{\cal O}^{(\tau)}_{I,\ell}} \sim \frac{\ell^{\tau-3/2}}{4^\ell} \\
& &\sum_I c^{(0)}_{qp{\cal O}^{(p+q-\tau)}_{I,\ell}} c^{(0)}_{pq{\cal O}^{(p+q-\tau)}_{I,\ell}}  \sim \frac{\ell^{p+q-\tau-3/2}}{4^\ell} 
\end{eqnarray}
The correct divergence in perturbation theory implies a similar behaviour for the quantum OPE coefficients, so that we define

\begin{eqnarray}
& &\sum_I  c_{pp{\cal O}^{(\tau)}_{I,\ell}} c_{qq{\cal O}^{(\tau)}_{I,\ell}} =\frac{\ell^{\tau-3/2}}{4^\ell} \hat a^{(\tau)}(\ell) \\
& &\sum_I c_{qp{\cal O}^{(p+q-\tau)}_{I,\ell}} c_{pq{\cal O}^{(p+q-\tau)}_{I,\ell}}  =\frac{\ell^{p+q-\tau-3/2}}{4^\ell} \hat a^{(p+q-\tau)}(\ell)
\end{eqnarray}
Since in general the intermediate operators on both sides  of (\ref{crossingpqt}) are degenerate at tree level, we introduce the following weighted averages on each side

\begin{equation}
\langle f(\ell) \rangle_L = \frac{\sum_I  c_{pp{\cal O}^{(\tau)}_{I,\ell}} c_{qq{\cal O}^{(\tau)}_{I,\ell}} f_{I}(\ell)}{\sum_I  c_{pp{\cal O}^{(\tau)}_{I,\ell}} c_{qq{\cal O}^{(\tau)}_{I,\ell}} },~~~~~\langle f(\ell) \rangle_R = \frac{ \sum_{\ell,I} c_{qp{\cal O}^{(p+q-\tau)}_{I,\ell}} c_{pq{\cal O}^{(p+q-\tau)}_{I,\ell}}  f_{I}(\ell) }{ \sum_{\ell,I} c_{qp{\cal O}^{(p+q-\tau)}_{I,\ell}} c_{pq{\cal O}^{(p+q-\tau)}_{I,\ell}} }
\end{equation}
Note that the two averages are with respect to different weights. Using the results of appendix \ref{divappendix} we arrive to the following integral relation

\begin{eqnarray}
\label{integralpqtau}
\frac{1}{\Gamma^2\left(\frac{\tau}{2} \right)} \int_0^\infty \hat a^{(\tau)} \left(\frac{x}{\sqrt{v}} \right)  \langle 2^{\gamma^{(\tau)} \left( \frac{x}{\sqrt{v}} \right)} u^{\frac{\gamma^{(\tau)} \left( \frac{x}{\sqrt{v}} \right)}{2}} \rangle_L x^{\tau-1} K_0(2x) dx=&\\
\frac{1}{\Gamma(p-\frac{\tau}{2})\Gamma(q-\frac{\tau}{2})} \int_0^\infty   \hat a^{(p+q-\tau)} \left( \frac{x}{\sqrt{u}} \right) \langle 2^{\gamma^{(p+q-2)} \left( \frac{x}{\sqrt{u}} \right)} v^{\frac{\gamma^{(p+q-2)} \left( \frac{x}{\sqrt{u}} \right)}{2}} \rangle_R& x^{p+q-\tau-1} K_{p-q}(2x) dx \nonumber
\end{eqnarray}
Two comments are in order. First, in the above equation we keep only terms that are not suppressed by powers of the spin in the large spin expansions of $\gamma(\ell)$ and $\hat a(\ell)$ on both sides. Second, we have chosen a normalization such that in perturbation theory 

\begin{equation}
 \hat a^{(\tau)}  = 1 + \cdots, ~~~~~\hat a^{(p+q-\tau)}  = 1 + \cdots
\end{equation}
Once a solution is found, we can always multiply both sides by a function of the coupling constant, and the resulting OPE coefficients will still be a solution. Relation (\ref{integralpqtau}) imposes non-trivial constraints on the spectrum and OPE coefficients. It turns out these constraints are simpler to analyse in the case in which the tower $HS_\tau$ is non-degenerate, namely $\tau=2$. In this case the equation reduces to

\begin{eqnarray}
\label{couplepq}
\int_0^\infty \hat a^{(2)} \left(\tfrac{x}{\sqrt{v}} \right) 2^{\gamma^{(2)}(\frac{x}{\sqrt{v}})} u^{\frac{\gamma^{(2)}\left( \frac{x}{\sqrt{v}}\right)}{2}}  x K_0(2x) dx=\\
\frac{1}{\Gamma(p-1)\Gamma(q-1)} \int_0^\infty x^{p+q-3}  &\hat a^{(p+q-2)} \left( \frac{x}{\sqrt{u}} \right) \langle 2^{\gamma^{(p+q-2)} \left( \frac{x}{\sqrt{u}} \right)} v^{\frac{\gamma^{(p+q-2)} \left( \frac{x}{\sqrt{u}} \right)}{2}} \rangle K_{p-q}(2x) dx \nonumber
\end{eqnarray}
Below we will explicitly consider the constraints arising from this relation and comment on the general case. Before that, however, let us comment on the other crossing relation. Compute at tree-level

\begin{equation}
G_{pqpq}(u,v) = u^{\frac{|q-p|}{2}} d_{00} + u^{\frac{|q-p|}{2}+1}\left(d_{10} + \frac{d_{11}}{v} \right) + \cdots = \sum_{j=0}^i \sum_{i=0}^{min(p,q)} \frac{u^{\frac{|q-p|}{2}+i}}{v^j} d_{ij}
\end{equation}
In the quantum theory $d_{ij} \to d_{ij}(\log u,\log v)$. Crossing implies

\begin{equation}
G_{pqpq}(u,v) = \frac{u^{\frac{p+q}{2}}}{v^\frac{p+q}{2}} \sum_{j=0}^i \sum_{i=0}^{min(p,q)} \frac{v^{\frac{|q-p|}{2}+i}}{u^j} d^c_{ij}
\end{equation}
So that we have a relation of the form 

$$HS_{|q-p|+2m} \leftrightarrow HS_{p+q-2m}$$
However, there is a crucial difference with the previous case. According to the results of appendix \ref{treecase}:

\begin{eqnarray}
\sum_I c_{pq\ell} c_{pq \ell} = \frac{\ell^{p+q-2m-3/2}}{4^\ell} (-1)^\ell \tilde \alpha_0  +  \frac{\ell^{2 min(p,q)-2m-3/2}}{4^\ell}\alpha_1  + \cdots
\end{eqnarray}
so that the leading divergence as $v \to 0$ does not arise from the leading behaviour of the OPE coefficients but rather from a subleading term, which does not contain $(-1)^\ell$. The consequences of this are that if we were to define $\hat a(\ell)$ as above, not only the leading term would contribute, but also terms which are suppressed in the large spin limit, provided they contain an additional $(-1)^\ell$. The same will happen with the anomalous dimension contributions. For this reason, in the following we will focus on relations (\ref{integralpqtau}) and (\ref{couplepq}).
 
 \subsection{Solving the integral equation}
 
As we have seen, relation (\ref{integraltwist2}) implies a logarithmic behaviour for the anomalous dimension of twist-two operators and fixes completely the large spin behaviour of the OPE coefficients, both results valid to all loops in perturbation theory. In the following we would like to work out the implications of (\ref{couplepq}). 

First, note that at tree-level $\langle ... \rangle=1$, $\hat a = 1$, and all anomalous dimensions vanish so that the integral relation is satisfied. As we turn on the coupling it follows that the average anomalous dimensions for twist $p+q-2$ operators can have at most a logarithmic behaviour, very much as for the twist two case. So that

\begin{eqnarray}   
\label{gammaexp}
\langle \gamma^{(p+q-2)}(\ell) \rangle &=& \langle \rho \rangle \log \ell + \langle \beta \rangle+\cdots, \\
\langle \left(\gamma^{(p+q-2)}(\ell)\right)^2 \rangle &=& \langle \rho^2 \rangle \log^2 \ell  +2  \langle \rho \beta \rangle \log \ell+ \langle \beta^2 \rangle+\cdots, 
\end{eqnarray}
and so on. This is consistent with the analysis of \cite{Belitsky:2008mg}. Note that due to degeneracy in general $\langle \gamma^2 \rangle \neq \langle \gamma \rangle^2$. Each of the quantities of the r.h.s. will have a coupling constant dependence, so that
\begin{eqnarray}
\rho = \rho_1 g + \rho_2 g^2 + \cdots\\
\beta = \beta_1 g + \beta_2 g^2 + \cdots
\end{eqnarray}
This is to be supplemented with the known behaviour for the anomalous dimension of twist two operators. Crossing symmetry implies a similar logarithmic behaviour for the average of the OPE coefficients:

\begin{eqnarray}\label{conv}
\hat a^{(2)}(\ell) &=& 1 + g(a_{10} + a_{11} \log \ell) + g^2 (a_{20} + a_{21} \log \ell + a_{22} \log^2\ell) + \cdots \\
 \hat a^{(p+q-2)}(\ell) &=& 1 + g(a^{(pq)}_{10}+  a^{(pq)}_{11}  \log \ell) + g^2 ( a^{(pq)}_{20}  + a^{(pq)}_{21} \log \ell +  a^{(pq)}_{22} \log^2\ell) + \cdots 
\end{eqnarray}
We could insert all the corresponding expansions into (\ref{couplepq}), expand order by order in perturbation theory and work out the corresponding constraints. We can also proceed in a more systematic way. First rewrite the integral equation as

\begin{eqnarray}
\label{integralxy}
\int_0^\infty \hat a^{(2)} \left(\tfrac{x}{\sqrt{v}} \right) 2^{\gamma^{(2)}(\frac{x}{\sqrt{v}})}  u^{\frac{\gamma^{(2)}\left( \frac{x}{\sqrt{v}}\right)}{2}} x K_0(2x) dx=\\
\frac{1}{\Gamma(p-1)\Gamma(q-1)} \int_0^\infty y^{p+q-3}  &\hat a^{(p+q-2)} \left( \frac{y}{\sqrt{u}} \right) \langle 2^{\gamma^{(p+q-2)} \left( \frac{y}{\sqrt{u}} \right)} v^{\frac{\gamma^{(p+q-2)} \left( \frac{y}{\sqrt{u}} \right)}{2}} \rangle K_{p-q}(2y) dy \nonumber
\end{eqnarray}
with the following logarithmic behaviour for the anomalous dimensions:

\begin{eqnarray}
\gamma^{(2)}(\ell)&=& g \log \ell+b,\\
  \langle (\gamma^{(p+q-2)})^n \rangle& =& \langle (\rho \log \ell+ \beta)^n \rangle. 
\end{eqnarray}
Then introduce the following integral representations:
\begin{eqnarray}
\label{integralrep}
\hat{a}^{(2)} \left(\tfrac{x}{\sqrt{v}} \right)  = \int  F^{(2)}(y,\tfrac{x}{\sqrt{v}})K_{p-q}(2y) dy\\
\hat{a}^{(p+q-2)} \left( \tfrac{y}{\sqrt{u}} \right)= \int F^{(pq)}(x, \tfrac{y}{\sqrt{u}} ) K_0(2x) dx \nonumber
\end{eqnarray}
Plugging this into (\ref{integralxy}) we obtain an equation of the form 

\begin{equation}
\int P_L(x,y,u,v) K_0(2x)K_{p-q}(2y)dx dy = \int P_R(x,y,u,v)  K_0(2x)K_{p-q}(2y)dx dy
\end{equation}
where $P_L$ and $P_R$ have a very specific form. It turns out that to any order in perturbation theory the Kernel $K_0(2x)K_{p-q}(2y)$ is such that the above equation actually implies $P_L=P_R$. It turns out this implies the following remarkable property for the average of the spectrum:

\begin{equation}
\langle \rho^n \rangle =\langle \rho \rangle^n,~~~ \langle \rho^m \beta^n \rangle = \langle \rho \rangle^m \langle \beta^n \rangle
\end{equation}
and furthermore
\begin{equation}
\langle \rho \rangle =g 
\end{equation}
So that to any order in perturbation theory the leading logarithmic behaviour of the averaged anomalous dimension of the $p+q-2$ higher spin operator behaves as if there were no degeneracy and equals the anomalous dimension of twist two operators! This is not in contradiction with \cite{Belitsky:2008mg}, since here we are only talking about a weighted average and in the limit of large spin. Furthermore, crossing also fixes

\begin{eqnarray}
F^{(2)}(x,\zeta) &=&  \alpha 2^{-g \log \zeta -b+\beta } x^{p+q-3-b-g \log \zeta} \zeta^{-\beta} \\
F^{(pq)}(x,\zeta)&=&\alpha 2^{-g \log \zeta} x^{1-\beta-g \log \zeta} \zeta^{-b}
\end{eqnarray}
with the understanding that powers of $\beta$ are to be understood in an averaged sense \footnote{For instance $2^\beta \to \langle 2^\beta \rangle = 1 + \langle \beta \rangle \log 2+\cdots$ }. $\alpha$ is an arbitrary function of the coupling constant (not fixed by crossing) but independent of the spin. Plugging this back into the integral representations (\ref{integralrep}) we obtain

\begin{eqnarray}
\hat{a}^{(2)}(\ell) &=& \alpha(g) 2^{-g \log \zeta -b+\beta} \ell^{-\beta} \Gamma\left(p-1 - \frac{1}{2}  \gamma^{(2)}(\ell) \right)\Gamma\left(q-1 - \frac{1}{2}  \gamma^{(2)}(\ell) \right) \label{eqnfora2}\\
\hat{a}^{(p+q-2)}(\ell) &=& \alpha(g) 2^{-g \log \zeta} \ell^{-b} \Gamma\left(1 - \frac{1}{2}  \gamma^{(p+q-2)}(\ell) \right)^2 \label{eqnforapq}
\end{eqnarray}
where $ \gamma^{(p+q-2)}(\ell)= g \log \ell +\beta$ and again, powers of $\beta$ (which arise when expanding the expression above) are to be understood in a averaged sense. Two comments are in order. First recall $a^{(2)}(\ell)$ arose from a factorised OPE coefficient  $c_{pp{\cal O}^{(2)}_\ell}c_{qq{\cal O}^{(2)}_\ell}$, so that the factor $\langle \ell^{-\beta} \rangle$ should factorize accordingly, namely  $\langle \ell^{-\beta} \rangle = f(p) f(q)$. Furthermore note that the rest of the answer factorises as well. Second, note that from our answer we can read off the following universal behaviour at large spin 

\begin{equation} 
c_{pp{\cal O}^{(2)}_\ell} \sim c^{(0)}_{pp{\cal O}^{(2)}_\ell} \Gamma\left(p-1 - \frac{1}{2}  \gamma^{(2)}(\ell) \right)
\end{equation} 
Up to a prefactor which depends on the details of the theory. The result for $\hat a^{(p+q-2)}(\ell) $ has a similar universal behaviour (but in this case $\hat a^{(p+q-2)}(\ell)$ is itself a sum over many contributions), namely

\begin{equation}
c_{pq{\cal O}^{(p+q-2)}_\ell} \sim \Gamma\left(1 - \frac{1}{2}  \gamma^{(p+q-2)}(\ell) \right)
\end{equation} 
The universal behaviour we have found can be summarised as follows. The OPE coefficient between two scalar operators of weights $\Delta_p$ and $\Delta_q$ and a higher spin operator of (tree-level) twist $\tau$ has the universal behaviour

\begin{equation}
c_{pq {\cal O}^{(\tau)}_\ell} \sim \Gamma \left( \frac{\Delta_p + \Delta_q - \tau - \gamma_\ell^{(\tau)}}{2}\right)
\end{equation}
where averages should be understood where it corresponds. This behaviour is also consistent with the most general relation (\ref{integralpqtau}), but in this case the prefactor is more complicated. This structure is very reminiscent of the result one would obtain from Witten's diagrams in supergravity. However, in the present paper we have only analysed the consequences of crossing, without any further assumptions.  

\subsection{Comparison to polygonal Wilson loops}

\begin{figure}[h!] \centering
\begin{tikzpicture}[scale=3]
\coordinate [label=left:\textcolor{blue}{$A$}] (A) at (0,0);
\coordinate [label=right:\textcolor{blue}{$B$}] (B) at (1,0);
\coordinate [label=right:\textcolor{blue}{$C$}] (C) at (1,1);
\coordinate [label=left:\textcolor{blue}{$D$}] (D) at (0,1);
\draw [green,fill] (A) rectangle (C);

\draw [ultra thick,red] (0,0) -- (1,0); 
\draw [ultra thick,red] (0,0) -- (0,1);
\draw [ultra thick,red] (1,0) -- (1,1);
\draw [ultra thick,red] (0,1) -- (1,1);

\draw [fill] (0,0) circle [radius=0.02];
\draw [fill] (1,0) circle [radius=0.02];
\draw [fill] (0,1) circle [radius=0.02];
\draw [fill] (1,1) circle [radius=0.02];

\draw [thick, <-] (0,-1/6) -- (1/3,-1/6) node [right] {$\Delta \sigma$};
\draw [thick, ->] (2/3,-1/6) -- (1,-1/6) ;

\draw [thick, <-] (-1/4,0) -- (-1/4,1/3+1/15) node [above] {$\Delta \tau$};
\draw [thick, ->] (-1/4,2/3-1/15) -- (-1/4,1) ;

\end{tikzpicture}
\caption{As consecutive insertion points become null separated in space-time they form a rectangle $(A,B,C,D)$ in the $(\sigma,\tau)$ coordinates. \label{WLpicture}}
\end{figure}
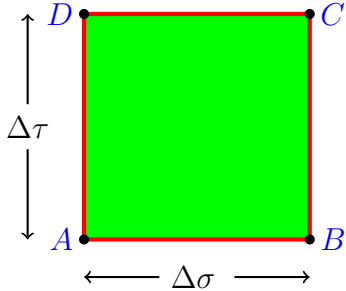

The consecutive null limit $x_{i,i+1}^2 \to 0$ of correlators in conformal gauge theories was studied in \cite{Alday:2010zy}. In this limit there are fast particles propagating between consecutive points and the correlator should reduce to the expectation value of a polygonal Wilson loop. For the particular case of a four-point function this limit coincides with the double null limit where $u,v \to 0$ at the same rate. It was argued in \cite{Alday:2010zy} (see section 4 of that paper) that in this limit we should obtain

\begin{equation}
\label{WL}
\lim_{u,v \to 0} \frac{G_{conn}}{ G_{conn}^{tree}} \sim e^{-\frac{\Gamma_{cusp}}{4} \log u \log v + \frac{b_1}{2} \log u + \frac{b_2}{2} \log v } J(u,v)
\end{equation}
where $G$ denotes the full correlator (not only its conformal invariant part) and we focus on a given connected contribution, so that the fast particles can frame the Wilson loop, and divide by the corresponding connected piece at tree level. The result (\ref{WL}) can be better understood by choosing coordinates where the insertion points are at the vertices of a large rectangle with sides $\Delta \tau \approx -1/2 \log u$ and $\Delta \sigma \approx -1/2 \log v$, see figure \ref{WLpicture}.

As we approach the double null limit we have a fast moving particle going between the different vertices of the polygon. Since this particle is coloured, it sources a colour electric field which is extended on the rectangle (green area in fig. \ref{WLpicture}). This colour electric flux has constant energy density in the $(\tau,\sigma)$ plane, proportional to $\Gamma_{cusp}$, and this produces the leading divergence in the exponential in (\ref{WL}), proportional to the area of the rectangle. In the interacting theory the particles can interact with the flux, and there are further terms. The simplest contribution arises due to corrections to the energies of the propagating particles. These corrections are confined to the edge of the rectangle (red lines in fig. \ref{WLpicture}) and give rise to the subleading divergences in the exponential in (\ref{WL}), proportional to the perimeter of rectangle. Finally, the factor $J(u,v)$ arises from the fact that the particles are coloured so we can have a back reaction of the colour electric field on the propagation of the particles. Although this is in general a complicated factor some features were studied in \cite{Alday:2010zy}.

In the present paper we have computed the structure constants of higher spin operators with leading twist. In particular, these operators dominate the correlator in the small $u,v$ limit. Consider the correlator $G_{ppqq}$ above. In the small $u,v$ limit, with their ration fixed, only the functions $c_{ii}$ will survive. Each function corresponds to a different connected contribution. In particular, let us consider $c_{11}$ which corresponds to

\begin{equation}
\langle {\cal A}^p(x_1) {\cal A}^p(x_2) {\cal A}^q(x_3) {\cal A}^q(x_4) \rangle = \frac{c_{11}(\log u,\log v)}{x_{12}^{2p-2} x_{23}^2 x_{34}^{2q-2} x_{41}^2} + \cdots
\end{equation}
Plugging (\ref{eqnfora2}) back into the l.h.s. of (\ref{couplepq}) we obtain

\begin{equation}
\label{doublenull}
c_{11} \sim e^{-\frac{g}{4} \log u \log v + \frac{b}{2} \log u + \frac{\beta}{2} \log v } J(u,v)
\end{equation}
Recall that in our conventions $\Gamma_{cusp} \equiv g$. Hence, our result exactly agrees with (\ref{WL})!  Furthermore note that from the point of view of the picture in  \cite{Alday:2010zy} $\beta$ in the exponent in (\ref{doublenull}) arises from corrections to the energy of the particle going from $A$ to $B$ plus corrections to the energy of the particle going from $D$ to $C$. For the present case the first correction should depend only on $p$, while the second should depend only on $q$, leading to a factorised dependence, in agreement with the comment below (\ref{eqnfora2}-\ref{eqnforapq}). Finally, crossing symmetry provides an all loop expression for the factor $J(u,v)$:

\begin{equation}
 J(u,v) = \int_0^\infty dx dy x^{1-\beta + \frac{g}{2} \log u} y^{p+q-3-b + \frac{g}{2} \log v} e^{-g \log x \log y} K_0(2x) K_{p-q}(2y)
\end{equation}
where powers of $\beta$ should be understood as averages. This answer satisfies all the general properties for $J(u,v)$ discussed in  \cite{Alday:2010zy}. Other terms $c_{ii}$ will behave similarly . Hence, our results are in perfect agreement with the correspondence proposed in \cite{Alday:2010zy}.

\section{A case with global charge: ${\cal N}=4$ SYM} 
\label{MSYM}
In the simple model studied above we have ignored two important features. On one hand, a generic gauge CFT contains fermions and gauge bosons in addition to scalars. As a result, there will be higher spin operators also formed by non-scalar letters. In particular, some of these will have the same quantum numbers as the higher spin operators studied above, increasing the degeneracy. On the other hand, gauge theories can posses global symmetries, such that scalars and fermions are charged under this global symmetry. Projecting in different representations may split a priori degenerate higher spin operators. In the following we will see how the picture introduced in sections 2 and 3 works for the particular example of ${\cal N}=4$ SYM. 
\subsection{Higher spin towers in ${\cal N}=4$ SYM}

Four-dimensional ${\cal N}=4$ SYM posses a global $SU(4)_R$ $R-$symmetry group. Gauge invariant operators are formed by traces of the fundamental fields of the theory:  Scalar fields $\varphi^I$ in the ${\bf 6}$ of $SU(4)_R$; fermionic fields $\lambda^{A}_\alpha$ and $\bar \lambda_{A \dot \alpha}$, in the ${\bf 4}$  and ${\bf \bar 4}$ of $SU(4)_R$ and gauge bosons $A_\mu$ in the singlet representation; together with their derivatives. We can form the following higher spin operators of spin $\ell$, classified by their twist and $SU(4)$ representation:

\subsubsection*{Twist 2}
\begin{itemize}
\item $Tr \varphi \partial_{\mu_1} \cdots \partial_{\mu_\ell} \varphi$, transforming in the ${\bf 6} \times {\bf 6} = {\bf 1} + {\bf 15} + {\bf 20}'$.\\
\item $Tr \bar \lambda \Gamma_\mu \partial_{\mu_2} \cdots \partial_{\mu_\ell} \lambda$, transforming in the ${\bf 4} \times {\bf \bar 4} = {\bf 1} + {\bf 15}$.\\
\item $Tr F_{\nu \mu_1} \partial_{\mu_2} \cdots \partial_{\mu_{\ell-1}} F_{\mu_\ell}^\nu$, transforming in the ${\bf 1} \times {\bf 1} = {\bf 1}$
\end{itemize}
As a result, we can consider different towers $HS_{2}^{\cal R}$. Note that $HS_{2}^{\bf 20'}$ can only be formed by scalars so that it is still non-degenerate. 

\subsubsection*{Twist 3}
\begin{itemize}
\item $Tr \varphi  \varphi \partial_{\mu_1} \cdots \partial_{\mu_\ell} \varphi$, transforming in the $ {\bf 6} \times {\bf 6} \times {\bf 6} = 3 \times {\bf 6} + {\bf 10} + {\bf \bar 10} + {\bf 50} + 2 \times {\bf 64}$.\\
\item $Tr \varphi  \bar \lambda \Gamma_\mu \partial_{\mu_2} \cdots \partial_{\mu_\ell} \lambda$, transforming in the $ {\bf 6} \times {\bf 4} \times {\bf \bar 4} = 2 \times {\bf 6} + {\bf 10} + {\bf \bar 10} + {\bf 64}$.\\
\item $Tr \varphi  F_{\nu \mu_1} \partial_{\mu_2} \cdots \partial_{\mu_{\ell-1}} F_{\mu_\ell}^\nu$, transforming in the $ {\bf 6}  \times {\bf 1} \times {\bf 1} = {\bf 6}$
\end{itemize}
Again, note that $HS_{3}^{\bf 50}$ can only be made by scalars. 

\subsubsection*{Twist p}
A similar analysis can be performed for higher and higher twists, with a richer and richer structure. The upshot is that the representation with Dynkin labels $[0p0]$ and hence $HS_p^{[0p0]}$ can only be obtained from scalars. In particular ${\bf 20'}$ and ${\bf 50}$ correspond to $[020]$ and $[030]$ respectively. 

\subsection{Crossing symmetry in ${\cal N}=4$ SYM }

In four dimensional $\mathcal{N}=4$ SYM there is a class of half-BPS superconformal primary operators,  transforming in the $[0,p,0]$ of $SU(4)_R$ and with protected dimension $\Delta=p$. They are given by
\begin{equation}
O^{[p]}(x,t)= t_{r_1}\dots t_{r_p} Tr(\varphi^{r_1}\dots \varphi^{r_p})(x)
\end{equation}
where $r_i=1,\cdots,6$ and $t$ is a complex six dimensional null vector which encodes the R-symmetry structure. Superconformal symmetry fixes the structure of the four point function of such operators to be of the form \cite{Nirschl:2004pa}
\begin{multline} \label{fourpoint}
\langle O^{[p_1]}(x_1,t_1) O^{[p_2]}(x_2,t_2) O^{[p_3]}(x_3,t_3) O^{[p_4]}(x_4,t_4)\rangle\\
=\frac{\left( t_1 \cdot  t_2\right)^{p_1+p_2} \left( t_3 \cdot  t_4\right)^{p_3+p_4}}{x_{12}^{p_1+p_2} x_{34}^{p_3+p_4}}\left(\frac{x_{24} t_1 \cdot t_4}{x_{14} t_2 \cdot t_4} \right)^{p_1-p_2} \left(\frac{x_{14} t_1 \cdot t_3}{x_{13} t_1 \cdot t_4} \right)^{p_3-p_4} \mathcal{G}^{[p_1 p_2 p_3 p_4]}(u,v,\sigma, \tau)
\end{multline}
where we have introduced harmonic cross ratios $\sigma$ and $\tau$ defined as
\begin{equation}
\sigma=\frac{t_1 \cdot t_2\, t_3 \cdot t_4}{t_1 \cdot t_3\, t_2 \cdot t_4}=\alpha \bar{\alpha}\quad \quad \tau=\frac{t_1 \cdot t_4\, t_2 \cdot t_3}{t_1 \cdot t_3\, t_2 \cdot t_4} =(1-\alpha)(1-\bar{\alpha})
\end{equation}
Such correlator can be decomposed into $\frac{(p_1+1)(p_1+2)}{2}$ terms, accordingly to the different $SU(4)_R$ representation present in the OPE of $[0,p_1,0] \times [0,p_2,0]$ $\subset$ $[0,p_3,0] \times [0,p_4,0]$, where without loss of generality we assume that $p_1\leq p_2 \leq p_3 \leq p_4$. Each contribution, labelled by $(n,m)=[n-m, p_2-p_1+2m, n-m]$, may be decomposed in conformal blocks as
\begin{align} \label{blockdec}
 \mathcal{G}^{[p_1 p_2 p_3 p_4 ]}(u,v,\sigma, \tau)&=\sum_{0 \leq m \leq n \leq p_1} a^{(p_{12},p_{34})}_{nm}(u,v) Y^{(p_{12},p_{34})}_{n m}(\sigma, \tau)\\
a^{(p_{12},p_{34})}_{nm}(u,v) &= \sum_{\Delta, \ell} c^{(p_1,p_2)}_{n m, \Delta \ell}  c^{(p_3, p_4)}_{n m, \Delta \ell} g^{(p_{12},p_{34})}_{\Delta, \ell}(u,v)
\end{align}
where $Y^{(a,b)}_{n m}(\sigma, \tau)$ is written in terms of Jacobi polynomials $P^{(a,b)}_n(x)$ as
\begin{align}
Y^{(p_{12},p_{34})}_{n m}(\sigma, \tau)&=-\frac{(\alpha \bar{\alpha})^{p_{34}/2+1}}{2(\alpha-\bar{\alpha})}\left( P_{n+1}^{(\frac{p_{12}-p_{34}}{2},-\frac{p_{12}-p_{34}}{2})}(\frac{2}{\alpha}-1)P_{m}^{(\frac{p_{12}-p_{34}}{2},-\frac{p_{12}-p_{34}}{2}}(\frac{2}{\bar{\alpha}}-1)\right.\nonumber \\ 
& \left. +P_{m+1}^{(\frac{p_{12}-p_{34}}{2},-\frac{p_{12}-p_{34}}{2})}(\frac{2}{\alpha}-1)P_{n}^{(\frac{p_{12}-p_{34}}{2},-\frac{p_{12}-p_{34}}{2})}(\frac{2}{\bar{\alpha}}-1)\right)
\end{align}
Moreover superconformal Ward identities imply \cite{Chicherin:2015edu}
\begin{align}
 \mathcal{G}^{[p_1 p_2 p_3 p_4]}(u,v,\sigma, \tau)&=\mathcal{G}_{tree}^{[p_1 p_2 p_3 p_4]}(u,v,\sigma, \tau)+ \mathcal{F}^{[p_1 p_2 p_3 p_4]}(u,v,\sigma, \tau)\mathcal{G}_{loop}^{[p_1 p_2 p_3 p_4]}(u,v,g_{YM})
 \end{align}
where the function $\mathcal{G}_{loop}^{[p_1 p_2 p_3 p_4]}(u,v,g_{YM})$ admits a perturbative expansion in powers of $g_{YM}$. Let us stress that this factorised structure persists at any loop order.

Invariance of the four point function under the exchange $(x_1, t_1, p_1) \leftrightarrow (x_3, t_3, p_3)$ relates $\mathcal{G}^{[p_1 p_2 p_3 p_4]}(u,v,\sigma, \tau)$ to $\mathcal{G}^{[p_3 p_2 p_1 p_4]}(v,u,\tau, \sigma)$ through
\begin{equation} \label{crossingn4}
\mathcal{G}^{[p_1 p_2 p_3 p_4]}(u,v,\sigma, \tau)=\frac{u^{\frac{p_1+p_2}{2}} \tau^{\frac{p_2+p_3}{2}}}{v^{\frac{p_2+p_3}{2}}\sigma^{\frac{p_1+p_2}{2}}}\mathcal{G}^{[p_3 p_2 p_1 p_4]}(v,u,\tau, \sigma)
\end{equation}
It is easy to see from \eqref{crossingn4} that different $SU(4)_R$ representations will in general mix under crossing, and a given representation in the left hand side will map into a linear combination of all the possible representations on the right hand side. Notice however that the number of possible representations appearing on both sides of \eqref{crossingn4} is the same and given by $\frac{(p+1)(p+2)}{2}$ where $p$ is the smallest among the $p_i$. 

We would like to repeat the exercise of sections 2 and 3 for this case. In this setting we are considering the correlators $\mathcal{G}^{ppqq}$ and $\mathcal{G}^{qppq}$, where for the rest of the discussion we assume $p \leq q$. If we consider the small $u$ limit of $\mathcal{G}^{ppqq}(u,v,\sigma,\tau)$ the leading contribution (besides that of the identity operator) arises from twist two operators. As discussed above they can transform only in three $SU(4)_R$ representations, which in our conventions we denote by $(0,0)$, $(1,1)$ and $(1,0)$. Furthermore, we would like to focus in the leading divergence as $v$ goes to zero. We obtain
\begin{align}
\mathcal{G}^{ppqq}(u,v,\sigma,\tau)|_{HS_2}&= \frac{u}{v}  \frac{\tau}{\sigma} c_1\\
&= a^{(0,0)}_{00}(u,v) Y^{(0,0)}_{0 0}(\sigma, \tau)+ a^{(0,0)}_{10}(u,v) Y^{(0,0)}_{1 0}(\sigma, \tau)+a^{(0,0)}_{11}(u,v) Y^{(0,0)}_{1 1}(\sigma, \tau) \nonumber
\end{align}
where $c_1$ depends on $p$ and $q$ and on the specific gauge group. The functions $a^{(0,0)}_{i j}(u,v)$ admit the following expansion
\begin{equation} \label{exp}
a^{(0,0)}_{i j}(u,v)= \frac{u}{v} \left( c^{tree}_{i j}+ c^{loop}_{i j} f(\log u, \log v, g_{YM}) \right)
\end{equation}
and they can be expanded in collinear conformal blocks as in \eqref{blockdec}. As previously discussed, the representation $(1,1)$ is the only one which is non-degenerate and contains only operators built from scalars.\footnote{In principle the equation for representation different than $(1,1)$ are the same but harder to solve and they will involve weighted averages also on the direct channel.}
As a consequence of crossing symmetry we can write
\begin{align}
\mathcal{G}^{qppq}(u,v,\sigma,\tau)|_{HS_{p+q-2}}&=c_1 \frac{u^{p+q-2}}{v^{p-1}}  \frac{\tau^{p-1}}{\sigma^{\frac{p+q-2}{2}}}\\
&=\sum_{0 \leq m \leq n \leq p-1} \tilde{a}^{(q-p,p-q)}_{nm}(u,v) Y^{(q-p,p-q)}_{n m}(\sigma, \tau) \nonumber
\end{align}
The functions $ \tilde{a}^{(q-p,p-q)}_{nm}(u,v)$ admit the expansion
\begin{equation} \label{exp}
\tilde{a}^{(q-p,p-q)}_{i j}(u,v)= \frac{u^{p+q-2}}{v^{p-1}} \kappa \left( c^{tree}_{i j}+c^{loop}_{i j} f(\log v, \log u, g_{YM}) \right)
\end{equation}
where $\kappa$ is a coefficient which depends on $p$ and $q$. At this point it is clear that we can apply the same procedure and results of the previous sections provided we project in specific $SU(4)_R$ representations, namely 
$$HS^{(1,1)}_2 \leftrightarrow HS^{\mathcal{R}}_{p+q-2}$$
where $\mathcal{R}=[n-m, q-p+2m,n-m]$, for $0 \leq m \leq n \leq p-1$. So that for each of these representations we obtain a relation exactly as (\ref{couplepq}).

In \cite{Chicherin:2015edu}, the four point function of half-BPS operators of arbitrary dimensions have been computed in planar $\mathcal{N}=4$ SYM up to three loops. The simplest example with $p \neq q$ corresponds to the correlator $\mathcal{G}^{2233}(u,v,\sigma,\tau)$, so lets list the results for this case. By projecting this four point function in the {\bf{20'}} representation, one can perform the conformal partial wave expansion and extract $c_{22 {\cal O}^{(1,1)}_\ell}c_{33 {\cal O}^{(1,1)}_\ell}$ as well as $\gamma^{(1,1)}(\ell)$, up to two loops in perturbation theory. Using the notation of \eqref{conv} we obtain 
\begin{align}
a_{11}&=-\log 2\\
a_{22}&=\frac{1}{32} \left(\pi^2 + 16\log^2 2\right)
\end{align}
where $g$ is related to $a=\frac{g^2_{YM}N}{4 \pi^2}$ as
\begin{equation}
a=\frac{1}{2}g+\frac{1}{48}(24+\pi^2)g^2 + \cdots
\end{equation}
Now using the crossing relations (\ref{couplepq}) it is possible to compute the coefficients $\langle \beta \rangle$ and $\langle \beta^2 \rangle$ appearing in the expansions (\ref{gammaexp}) of the weighted averages of anomalous dimension of twist-3 operators for any of the three possible representation of $SU(4)_R$: 
\begin{align}
\langle \beta \rangle&=\left( -\frac{1}{2} + \gamma_e \right) g + \cdots\\
\langle \beta^2 \rangle&=\left( \frac{1}{2} - \gamma_e + \gamma_e^2 - \frac{\pi^2}{48} \right) g^2 + \cdots
\end{align}
where $\gamma_e$ is Euler Gamma constant.  As expected from unitarity, $\langle \beta^2 \rangle > \langle \beta \rangle^2$. 

We can also obtain results for general $p,q$. As noticed in \cite{Dolan:2004iy,Chicherin:2015edu}, three point functions of $c_{p_1 p_2 \ell}$ properly normalised are all equal at one loop since there is only one structure at this loop order. This allows computing $\langle \beta \rangle$ for generic $p,q$ giving
\begin{equation}
\label{sigmaMSYM}
\langle \beta \rangle=-\frac{1}{2}\left( \psi_0 \left( p-1\right)+ \psi_0\left(q-1\right) \right) g + \cdots
\end{equation}
where $\psi_0$ denotes the digamma function. Note that for $p=q=2$ this agrees with the finite piece of the anomalous dimension of twist two operators in the large spin limit, while for $p=2,q=3$ it agrees with the result given above. Furthermore, it displays the factorised structure discussed section 3.  
   
\section{Conclusions}
In the present paper we have studied weakly coupled conformal gauge theories by analytic bootstrap techniques. Weakly coupled gauge theories contain towers of higher spin operators of approximate twist $\tau$. By studying crossing symmetry for mixed correlators we have found that these towers (for large values of the spin) map to each other: 

$$HS_\tau \leftrightarrow HS_{\tau'}$$
This relation takes the form of an integral equation involving the spectrum and structure constants of the higher spin operators. In case of twists higher than two, such operators are degenerate and the integral relation involves weighted averages. Regarding the spectrum, we have found that crossing symmetry is consistent with a logarithmic behaviour, in agreement with \cite{Belitsky:2008mg}. Regarding the structure constants our results take the form

$$c_{pq {\cal O}_\ell^{(\tau)}}= f^{(\tau)}_{pq}(\ell) \times \Gamma \left( \frac{\Delta_p + \Delta_q -\tau -\gamma_\ell^{(\tau)}}{2} \right)$$
Namely, a universal factor times a theory-dependent prefactor $f^{(\tau)}_{pq}(\ell)$. The universal factor has a very similar structure to the one that arises when studying Witten diagrams. In particular, it includes a series of poles that start when the full twist of the higher spin operator equals the sum of the dimensions of the other two. In the context of large $N$ MSYM the appearance of analogous poles was analysed in \cite{Minahan:2014usa,Korchemsky:2015cyx} and where it was shown to be related to operator mixing. Although our results are in principle only valid in perturbation theory (but to all loops), given the discussions  in \cite{Minahan:2014usa,Korchemsky:2015cyx} we expect this structure to persist for finite $\gamma_\ell$, at least in the planar limit. It is very interesting this structure arises naturally by only requiring crossing symmetry. The theory-dependent prefactor, of the schematic form $f=\ell^{-\beta}$, depends on the theory under consideration and on averages that are hard to calculate. For the simplest case this prefactor is basically $f=\ell^{-b}$, where $b$ is the sub-leading/finite contribution to the anomalous dimension of twist two operators. In this case it does not add any new analytic structure to the answer, and we expect this to be the case in general.

Having solved for the constraints above one can then compute the mixed correlators under consideration in the double null limit. This limit was studied in  \cite{Alday:2010zy} where it was shown that the expectation value of a polygonal Wilson loop should be recovered. Our results are in perfect agreement with these expectation and furthermore they provide all loop results for certain prefactors that are in general hard to compute. 

Some open problems which we consider interesting are the following. The present paper generalises the results of \cite{Alday:2013cwa} to external operators with arbitrary dimension. This opens up the possibility to compare our results with results at strong coupling, since now we can consider $\Delta_p$ and $\Delta_q$ large. It would be very interesting to make a detailed comparison to the results of \cite{Minahan:2014usa} from string vertices. This may also allow to get a handle on the prefactor $f^{(\tau)}_{pq}(\ell)$ at strong coupling, ideally to compute it exactly. Regarding this, note that for ${\cal N}=4 SYM$ the explicit one-loop result (\ref{sigmaMSYM}) grows logarithmically as $p$ or $q$ becomes large. As a result $\ell^\beta$ becomes symmetric under $p \leftrightarrow \ell$. It would be interesting to understand this result.  

It would be interesting to explore further the relation to Wilson loops and the picture of \cite{Alday:2010zy}. The present paper offers a proof of the correlators/Wilson loop correspondence from crossing symmetry, for the four-dimensional case and gives explicit expressions for all ingredients involved. Can we learn more from this interplay? A related question is to understand our results, and in particular the structure of poles in the universal factor, along the lines of \cite{Alday:2007mf,Alday:2010zy}. This may provide a {\it finite coupling} understanding of the universal factor. 

Over the last years there has been progress in the computation of structure constants in planar ${\cal N}=4 SYM$ by integrability techniques. See for instance \cite{Basso:2015zoa} for the state of the art. Despite these developments, there are still missing ingredients if one wants to pursue the program to all loops. The present results may be useful in such endeavours. On one hand, the structures found in this paper should be visible in other approaches. Furthermore, the fact that the results of this paper are valid for any length of the external operators means that certain subtleties, such as wrapping, can be pushed away. 

It would also be interesting to apply these techniques to other weakly coupled gauge conformal field theories. An interesting example would be $\beta-$deformed ${\cal N}=4$ SYM. 

Finally, for theories with gravity dual (known or unknown) an interesting question is how much of the structure of the gravity dual can be understood from symmetries of CFT correlators. Or conversely, which CFT theories can admit a gravity dual. There has been a lot of activity in this regard, see for instance \cite{Heemskerk:2009pn} for early results in this direction and \cite{Perlmutter:2016pkf} for a different approach. It is remarkable that our results reproduce the pole structure of Witten diagrams. One may wonder if this would lead to a way to define constructively the would be gravity dual of our CFT's.    

\section*{Acknowledgments}

We are grateful to J. Maldacena, J. Minahan, E. Perlmutter and S. Zhiboedov  for useful discussions. A.B. acknowledges the University of Oxford for hospitality where part of this work has been done. The work of L.F.A was supported by ERC STG grant 306260. L.F.A. is a Wolfson Royal Society Research Merit Award holder. The work of A.B. is partially supported by Templeton Award 52476 of A. Strominger and by Simons Investigator Award from the Simons Foundation of X. Yin.

\appendix

\section{Divergent contributions from HS towers}
\label{divappendix}

In this appendix we present the leading divergence, as $v \to 0$, due to the exchange of higher spin operators in the direct channel, for various situations that we describe. The results below are heavily used in the body of the paper. The small $u$ limit of the scalar conformal block is given by, see {\it e.g.} \cite{Dolan:2011dv}.

\begin{eqnarray}
g_{\Delta,\ell}^{\Delta_{ij},\Delta_{kl}}(u,v) = u^{\frac{\Delta-\ell}{2}} f_{coll \,(\Delta,\ell)}^{\Delta_{ij},\Delta_{kl}}(v)
\end{eqnarray}
where the collinear part of the conformal block is given by

$$f_{coll \,(\Delta,\ell)}^{\Delta_{ij},\Delta_{kl}}(v)  =  (1-v)^\ell _2F_1\left( \frac{1}{2}(\Delta+\ell) -\frac{1}{2}\Delta_{ij}, \frac{1}{2}(\Delta+\ell) +\frac{1}{2}\Delta_{kl}, \Delta+\ell;1-v\right) $$
This result holds in general dimensions. Note that we are using conventions where we do not include an extra $(-1)^\ell$ factor in the conformal block. We are interested in computing the divergent contribution as $v \to 0$ of the following sum

\begin{eqnarray}
\sum_{\ell} a_\ell f_{coll \,(\Delta,\ell)}^{\Delta_{ij},\Delta_{kl}}(v)
\end{eqnarray}
where $\Delta = \Delta_0+\ell+\gamma_\ell$ and 

$$a_\ell = \frac{\ell^\kappa}{4^\ell} + \cdots$$
As discussed in \cite{Alday:2010zy,Komargodski:2012ek,Fitzpatrick:2012yx} the divergence arises from the large $\ell$ region and can be captured by focusing in the small $v$/large $\ell$ region. More precisely, we take $v \to 0$ keeping $x= \ell \sqrt{v}$ fixed. In this limit the sum over $\ell$ becomes an integral over $x$ and  we obtain

\begin{eqnarray}
\label{divergence}
\sum_{\ell \in 2 \mathbb{Z}} \frac{\ell^\kappa}{4^\ell} f_{coll\,(\Delta_0+\gamma_\ell +\ell,\ell)}^{\Delta_{ij},\Delta_{kl}}(v) = \frac{1}{v^{(3+2\kappa - \Delta_{ij}+ \Delta_{kl})/4}} \int_0^\infty dx x^{\kappa+\frac{1}{2}} \frac{2^{\Delta_0+\gamma}}{\sqrt{\pi}} K_{\frac{\Delta_{kl}- \Delta_{ij}}{2}}(2x) + \cdots
\end{eqnarray}
In the above expression we have assumed the sum runs over even spins only, which is the case, for instance, if we have identical external operators. In general we can have a sum over all spins. In this case:

 \begin{eqnarray}
\sum_{\ell \in \mathbb{Z}} \frac{\ell^\kappa}{4^\ell} f_{coll\,(\Delta_0+\gamma_\ell +\ell,\ell)}^{\Delta_{ij},\Delta_{kl}}(v) = \frac{2}{v^{(3+2\kappa - \Delta_{ij}+ \Delta_{kl})/4}} \int_0^\infty dx x^{\kappa+\frac{1}{2}} \frac{2^{\Delta_0+\gamma}}{\sqrt{\pi}} K_{\frac{\Delta_{kl}- \Delta_{ij}}{2}}(2x) + \cdots
\end{eqnarray}
In some cases odd spins contribute with a negative factor respect to even spins. In such case we do not get a divergent contribution. In other words

 \begin{eqnarray}
\sum_{\ell} (-1)^\ell \frac{\ell^\kappa}{4^\ell} f_{coll\,(\Delta_0+\gamma_\ell +\ell,\ell)}^{\Delta_{ij},\Delta_{kl}}(v) \sim 1
\end{eqnarray}
The results above are useful to compute the leading contribution from a given tower of higher spin operators. We may be interested in computing the divergent contribution due to descendants of these operators. In order to compute this we first need subleading corrections to collinear conformal blocks:

\begin{eqnarray}
g_{\Delta,\ell}^{\Delta_{ij},\Delta_{kl}}(u,v) = u^{\frac{\Delta-\ell}{2}} f_{coll \,(\Delta,\ell)}^{\Delta_{ij},\Delta_{kl}}(v) +u^{\frac{\Delta-\ell}{2}+1} f_{subcoll \,(\Delta,\ell)}^{\Delta_{ij},\Delta_{kl}}(v) + \cdots
\end{eqnarray}
These corrections have been computed in \cite{Alday:2015ewa} for identical external operators, in arbitrary dimensions, and for the particular case $d=4$ can be extracted from the known result for the scalar conformal blocks. The main result to be used in the body of the paper is that for all these cases the divergence due to descendants is of exactly the same order, namely

\begin{eqnarray}
\label{divergencedes}
\sum_{\ell } \frac{\ell^\kappa}{4^\ell} f_{subcoll\,(\Delta_0+\gamma_\ell +\ell,\ell)}^{\Delta_{ij},\Delta_{kl}}(v) \sim \frac{1}{v^{(3+2\kappa - \Delta_{ij}+ \Delta_{kl})/4}}
\end{eqnarray}
We expect this to be true for higher level descendants as well. 

\section{A tree-level case}
\label{treecase}

Consider correlators $G_{2323}(u,v)$ and $G_{2332}(u,v)$ at tree-level. Let us focus in the leading term, proportional to $u^{3/2}$, in the small $u$ expansion. One obtains

\begin{eqnarray}
G_{2323}(u,v) = u^{3/2} \left(\frac{a_0}{v} + a_1 \right)+ \cdots \\
G_{2332}(u,v) = u^{3/2} \left(\frac{b_0}{v^2} + \frac{b_1}{v} \right) + \cdots
\end{eqnarray}
In both cases, the divergences as $v \to 0$ arise as we sum over the tower of intermediate states $HS_3$. Note that the OPE coefficients entering in the expansions are related as $c_{32 \ell} = (-1)^\ell c_{23\ell}$. Furthermore, for intermediate states of twist three, the sum over spins runs over all natural numbers. At tree-level, we can assume an expansion of the form:

\begin{eqnarray}
\sum_I c_{23\ell} c_{23 \ell} = \frac{\ell^{\kappa_0}}{4^\ell}(\alpha_0 + (-1)^\ell \tilde \alpha_0 ) +  \frac{\ell^{\kappa_1}}{4^\ell}(\alpha_1 + (-1)^\ell \tilde \alpha_1 ) + \cdots \\
\sum_I c_{23\ell} c_{32 \ell}  = \frac{\ell^{\kappa_0}}{4^\ell}((-1)^\ell  \alpha_0 + \tilde \alpha_0 ) +  \frac{\ell^{\kappa_1}}{4^\ell}((-1)^\ell  \alpha_1 + \tilde \alpha_1 ) + \cdots
\end{eqnarray}
where $I$ runs over all operators for a given spin and $\kappa_0> \kappa_1$ will be fixed momentarily. In order to compute the divergent behaviour we use the results of appendix \ref{divappendix}. For the first correlator:

\begin{eqnarray}
\sum_{\ell} a_\ell \frac{\ell^\kappa}{4^\ell} f_{coll\,(\Delta_0 +\ell,\ell)}^{(-1,-1)}(v)=  \frac{\alpha_0}{v^{(3+2{\kappa_0})/4}} + \frac{\alpha_1}{v^{(3+2{\kappa_1})/4}} + \cdots 
\end{eqnarray}

While for the second correlator

\begin{eqnarray}
\sum_{\ell} (-1)^\ell a_\ell \frac{\ell^\kappa}{4^\ell} f_{coll\,(\Delta_0 +\ell,\ell)}^{-1,-1}(v) =  \frac{\tilde \alpha_0}{v^{(3+2\kappa_0+2)/4}} + \frac{\tilde \alpha_1}{v^{(3+2\kappa_1+2)/4}} + \cdots
\end{eqnarray}

The leading divergence of the second correlator implies $\kappa_0=\frac{3}{2}$ together with $\tilde \alpha_0\sim b_0$. Next, absence of a divergence $1/v^{3/2}$ in the first correlator implies $\alpha_0=0$. Then, the leading divergence of the first correlator implies $\kappa_1 = \frac{1}{2}$, together with $\alpha_1 \sim a_0$, while the second correlator implies $\tilde \alpha_1=0$. The conclusion of this discussion is that, at tree-level

\begin{eqnarray}
\sum_I c_{23\ell} c_{23 \ell} = \frac{\ell^{3/2}}{4^\ell} (-1)^\ell \tilde \alpha_0  +  \frac{\ell^{1/2}}{4^\ell}\alpha_1  + \cdots \\
\sum_I c_{23\ell} c_{32 \ell}  = \frac{\ell^{3/2}}{4^\ell} \tilde \alpha_0 +  \frac{\ell^{1/2}}{4^\ell}(-1)^\ell  \alpha_1+ \cdots
\end{eqnarray}
In the body of the paper we will be interested in a more general case, in which we consider correlators of the form $G_{pqpq}$ and $G_{qppq}$ and the contribution from $HS_{p+q-2m}$. The discussion proceeds exactly as above. At tree-level it is possible to compute 

\begin{eqnarray}
\left.G_{pqpq}\right|_{HS_{p+q-2m}} &=& u^{\frac{p+q-2m}{2}}\left( \frac{d}{v^{min(p,q)-m}}+\cdots \right)\\
\left.G_{qppq}\right|_{HS_{p+q-2m}} &=& u^{\frac{p+q-2m}{2}}\left( \frac{c}{v^{p-m}}+\cdots \right)
\end{eqnarray}
So that at tree-level

\begin{eqnarray}
\sum_I c_{pq\ell} c_{pq \ell} = \frac{\ell^{p+q-2m-3/2}}{4^\ell} (-1)^\ell \tilde \alpha_0  +  \frac{\ell^{2 min(p,q)-2m-3/2}}{4^\ell}\alpha_1  + \cdots \\
\sum_I c_{pq\ell} c_{qp \ell}  = \frac{\ell^{p+q-2m-3/2}}{4^\ell} \tilde \alpha_0 +  \frac{\ell^{2 min(p,q)-2m-3/2}}{4^\ell}(-1)^\ell  \alpha_1+ \cdots
\end{eqnarray}
where the intermediate operator has twist $p+q-2m$. For the case $p=3,q=2,\tau=2$ this reduces to the previous case. 

\section{Degeneracy of twist operators}
\label{degeneracyappendix}
In this appendix we study the degeneracy of primary operators with fixed twist, of the form 

\begin{equation}
Tr \varphi^i \partial^\ell \varphi^{L-i}
\end{equation}
where the derivative is along a fixed null direction. For such operators the twist coincides with the length $L$. The degeneracy of such operators can be easily computed by Polya theory. First, let us consider the single letter partition function:

\begin{equation}
Z_1(q) = q+ q^2 + q^3 + \cdots =\frac{q}{1-q}
\end{equation}
which counts states of the form $\partial^n \varphi$. The multi-letter partition function, taking into account cyclycity of the trace is given by

\begin{equation}
Z_L(q) =\frac{1}{L} \sum_{s=1}^L \left( Z_1(q^{\frac{L}{(s,L)}})  \right)^{(s,L)}
\end{equation}
where $(s,L)$ denotes the largest common divisor of $s$ and $L$. In order to compute the number of independent primaries, at each level we subtract the number of operators at previous level, so that

\begin{equation}
P_L(q) = (1-q)Z_L(q) 
\end{equation}
is the generating function for the number of primaries. For the first few twists we find

\begin{eqnarray}
P_2(q) &=& \frac{q^2}{1-q^2}\\
P_3(q) &=& \frac{q^3 ((q-1) q+1)}{(q-1)^2 \left(q^2+q+1\right)}
\end{eqnarray}
In particular primary operators with twist two are non-degenerate and have only even spin, while primary operators of twist three and higher are always degenerate. The degeneracy for large values of the spin can be understood from the behaviour near $q=1$. We find $d_L(\ell) \sim \ell^{L-2}$.

\end{document}